\documentclass[preprint]{elsarticle}

\usepackage{lineno,hyperref}
\usepackage{graphicx}
\usepackage{dcolumn}
\usepackage{bm}
\usepackage{subfigure}
\usepackage{braket}

\journal{Journal of Electron Spectroscopy and Related Phenomena}

\begin{document}

\begin{frontmatter}

\title{Advances in high-order harmonic generation sources for time-resolved investigations}


\author[1,2]{Maurizio~Reduzzi}
\author[1]{Paolo~Carpeggiani}
\author[3]{Sergei~K\"{u}hn}
\author[2]{Francesca~Calegari}
\author[1,2]{Mauro~Nisoli}
\author[1,2]{Salvatore~Stagira}
\author[2]{Caterina~Vozzi}
\author[3]{Peter~Dombi}
\author[3]{Subhendu~Kahaly}
\author[3,4]{Paris~Tzallas}
\author[3,5]{Katalin~Varju}
\author[3]{Karoly~Osvay}
\author[1,2,3]{Giuseppe~Sansone\corref{mycorrespondingauthor}}
\cortext[mycorrespondingauthor]{Corresponding author}
\ead{giuseppe.sansone@polimi.it; giuseppe.sansone@eli-alps.hu}

\address[1]{Dipartimento di Fisica, Politecnico di Milano Piazza Leonardo da Vinci 32, 20133 Milano Italy}
\address[2]{Institute of Photonics and Nanotechnologies, CNR-IFN, Piazza Leonardo da Vinci 32, 20133 Milano Italy}
\address[3]{ELI-ALPS, ELI-Hu Kft., Dugonics ter 13, H-6720 Szeged Hungary}
\address[4]{Foundation for Research and Technology-Hellas, Institute of Electronic Structure and Lasers B.O. Box 1527, GR-711 10 Heraklion Crete, Greece}
\address[5]{Department of Optics and Quantum Electronics, University of Szeged, D\'{o}m t\'{e}r 9, 6720 Szeged, Hungary}

\begin{abstract}
We review the main research directions ongoing in the development of high-harmonic generation-based extreme ultraviolet sources for the synthesization and application of trains and isolated attosecond pulses to time-resolved spectroscopy. A few experimental and theoretical works will be discussed in connection to well-established attosecond techniques. In this context, we present the unique possibilities offered for time-resolved investigations on the attosecond timescale by the new Extreme Light Infrastructure Attosecond Light Pulse Source, which is currently under construction.
\end{abstract}


\end{frontmatter}


\section{Introduction}
\label{sec1}
Since the first demonstration of high-order harmonic generation (HHG) in gases~\cite{JPB-Ferray-1988,PRL-Lhuillier-1993}, the efforts of several research groups, combined with the development of new technologies for the generation of intense, high-repetition rate driving sources in the near (IR) and mid-infrared (mid-IR) spectral range, has led to impressive progresses in the field of ultrafast extreme ultraviolet (XUV) spectroscopy and of attosecond science. After the first pioneering experiments in atoms~\cite{NATURE-Drescher-2002}, molecules~\cite{NATURE-Sansone-2010}, and condensed phase systems~\cite{NATURE-Cavalieri-2007}, time-resolved experiments using high-order harmonics are moving fast towards the investigation of more complex systems such as biomolecules~\cite{SCIENCE-Calegari-2014} and composite materials~\cite{NATURE-Neppl-2015}. All these experiments indicate that the fundamental steps of electronic dynamics evolve on the attosecond timescale, calling for the reproducible generation and characterization of sub-femtoseconds pulses to excite and probe electronic wave packets. While the generation of trains of attosecond pulses can be accomplished by using multi-cycle IR intense femtosecond pulses, the synthesization of isolated attosecond pulses requires the precise control of the electric field of the driving pulse~\cite{NATURE-Baltuska-2003} in combination with techniques for the (spectral or temporal) confinement of the harmonic generation mechanism~\cite{NATPHOT-Chini-2014}. Nowadays, pulse durations are quickly approaching the atomic unit of time (1~a.u.=24~as)~\cite{JPB-Sansone-2009, OL-Zhao-2012}.\par
Time-resolved studies based on high-order harmonic radiation, however, are affected by limitations that can be traced back to the fundamental characteristics of the HHG process. While the XUV-pump-IR-probe approach is routinely implemented in several laboratories, the conceptually more straightforward XUV-pump-XUV-probe approach still represents a formidable experimental challenge and it has been demonstrated only by a few groups worldwide~\cite{NATPHYS-Tzallas-2011,NATCOMM-Takahashi-2013}. The main reason resides in the low conversion efficiency of the HHG process (usually in the $10^{-9}-10^{-5}$ range) that calls for high-energy (several tens or hundreds of mJ) driving pulses for reaching the energy level required for multi-photon interaction and nonlinear absorption in the XUV spectral range.\par Similarly, HHG by driving pulses in the IR spectral range is limited to photon energies up to, typically, $\simeq$ 100 eV, thus preventing the possibility to address and investigate electronic dynamics initiated by excitation or ionization of core electrons. In this context the development of mid-IR driving source is motivated by the favorable scaling of the high harmonics cut-off energy on the wavelength of the driving pulse, which holds the promise to give access to keV isolated attosecond pulses~\cite{SCIENCE-Popmintchev-2012}.\par Finally, several experimental techniques, such as photoelectron microscopy, and coincidence photoelectron and photoion spectroscopy, require moderate pulse energy, but at (very) high repetition rates in order to overcome space-charge effects and to improve the signal-to-noise ratio. Recent technological developments indicate that generation of isolated attosecond pulses at MHz repetition rate should be within reach in the next few years~\cite{NATPHOT-Krebs-2013}.\par
The advent of Free Electron Lasers operating in the XUV/X-ray spectral range, however, has already allowed to overcome some of these limitations, and new opportunities for time-resolved studies will be offered by the continuous improvement of their characteristics (for example through the implementation of different seeding schemes that are expected to improve the shot-to-shot reproducibility and the longitudinal coherence~\cite{NATPHOT-Thompson-2010}). Apart from these new developments, the novel European Extreme Light Infrastructure Attosecond Light Pulse Source (ELI-ALPS) holds the promise to overcome several of the limitations currently affecting HHG-based sources, by offering users with an unprecedented combination of attosecond XUV sources in terms of pulse energy, photon spectral range and repetition rate. Moreover, the facility will offer the possibility to combine XUV radiation with pulses extending from the THz region up to the X-ray spectral range and with particle beams.\par
The manuscript is organized in the following way:\\
in section~\ref{sec2}, we provide an overview of the main development directions ongoing in laser technology for the generation of attosecond pulses at high-repetition rate ($>$~10~kHz)~(\ref{sec21}), with high energy per pulse (in the $\mathrm{\mu J}$ range)~(\ref{sec22}), and at high photon frequencies (hundreds of eV)~(\ref{sec23}). In section~\ref{sec3}, we review the main spectroscopic techniques implemented in the attosecond field, in connection with a few selected experimental and theoretical works. Finally, in the last section, we introduce the new possibilities offered by the new attosecond facility ELI-ALPS for the investigation of atomic, molecular and cluster dynamics under intense and ultrashort XUV light pulses.

\section{Current development directions in HHG}
\label{sec2}
\begin{figure}[htb]
\centering\includegraphics[width=10cm]{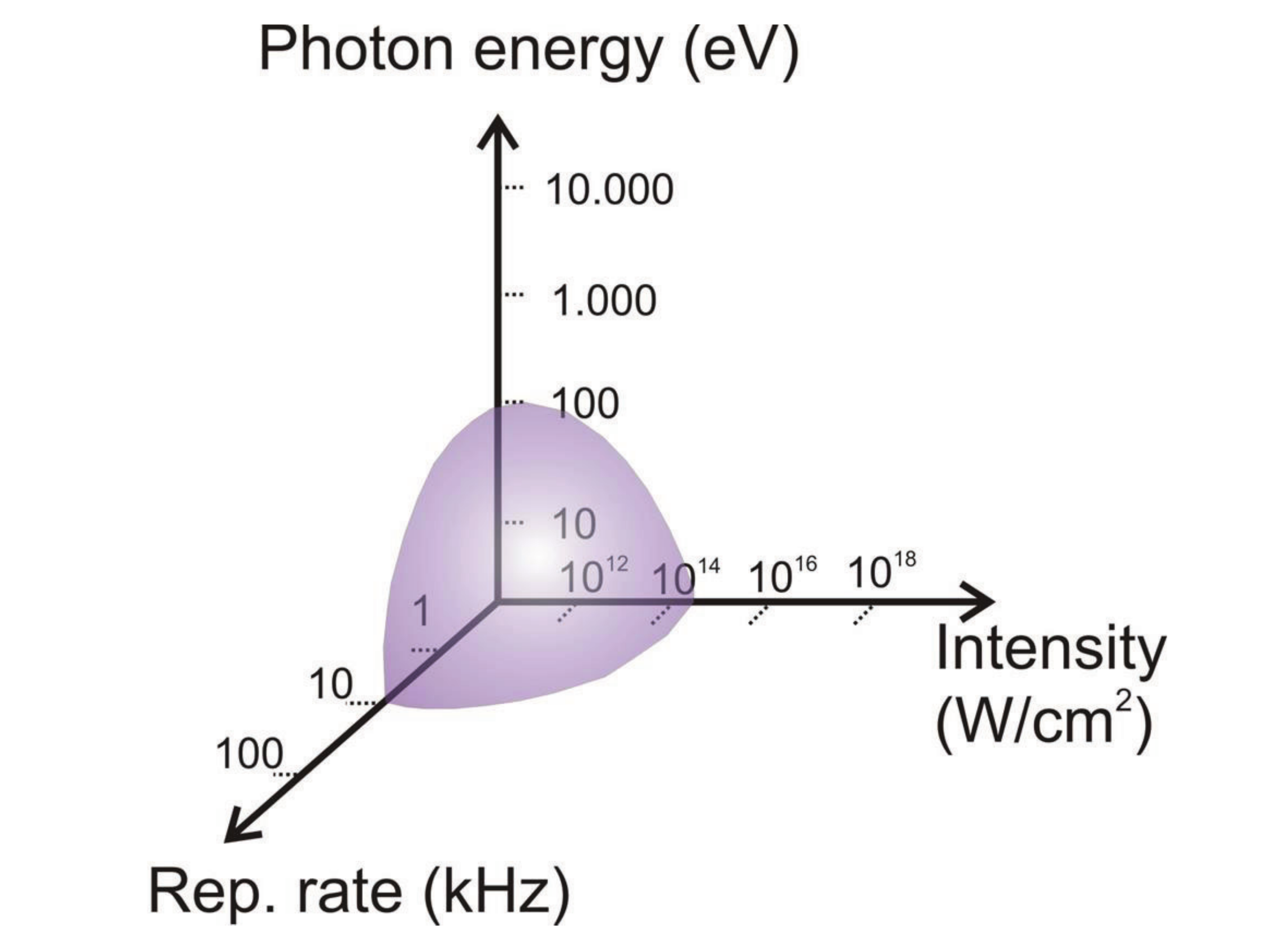}
\caption{The three main research directions in HHG sources are aiming for the development of high-repetition rate, high intensity, and high photon energy XUV pulses. The shaded area represents the parameter-space spanned by typical HHG-based XUV sources. State-of-the-art experimental setups will be discussed in the corresponding sections.}
\label{Fig1}
\end{figure}

Three main research directions can be identified in the development of HHG-based sources as schematically shown in Fig.~\ref{Fig1}: generation of attosecond pulses at high repetition rates (up to the MHz), generation of intense attosecond pulses for high-peak intensities, and extension of the cut-off energy in the keV range. These three directions call for different driving sources and will be discussed separately in the next sub-sections.

\subsection{HHG at high-repetition rates} \label{sec21}
The generation of high-order harmonics at high repetition rates (0.1 - 100 MHz) represents a milestone for atomic and molecular dynamics applications, such including photoion and photoelectron spectroscopy~\cite{IEEE-Sansone-2012} and for condensed phase and nanoplasmonics investigations using time-resolved photoelectron microscopy~\cite{RSI-Mikkelsen-2009}. Multi-kHz HHG was realized in 2003 using a titanium:sapphire (Ti:Sa) amplifier which delivered pulses with energies up to 7~$\mathrm{\mu J}$ at 100 kHz repetition rate \cite{PRA-Lindner-2003}, and ,recently, the first full temporal characterization of isolated attosecond pulse up to 10~kHz was reported~\cite{RSI-Sabbar-2014}.\par
\begin{figure*}[htb]
\centering\includegraphics[width=12cm]{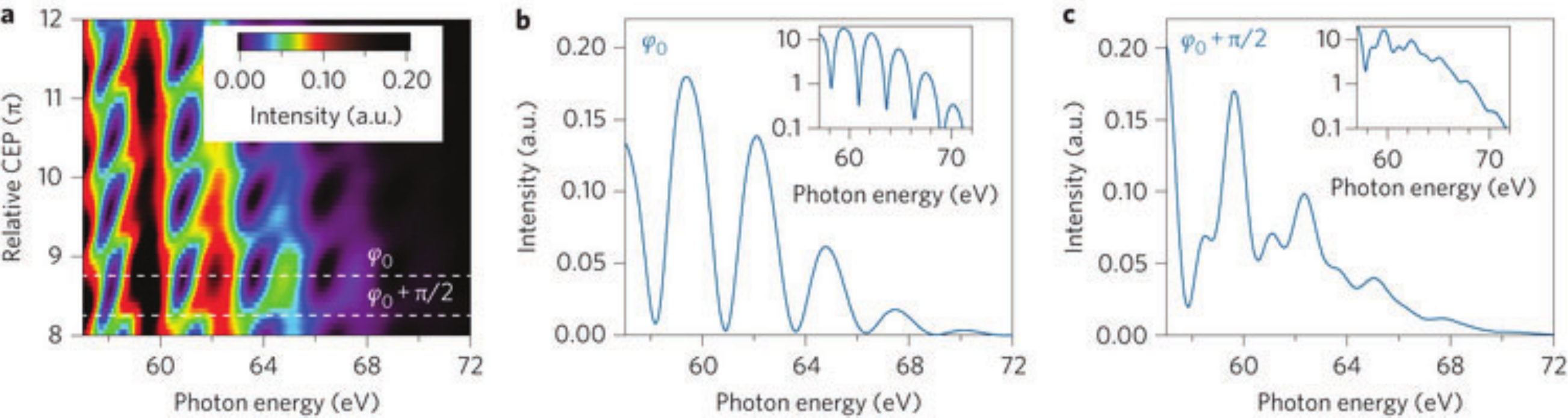}
\caption{a) XUV spectra generated by a few-optical-cycle system pumped by fiber laser operating at 600~kHz as a function of the CEP of the driving pulses. b) Spectral distribution for the CEP $\varphi=\varphi_0$ indicating the emission of, at least, two attosecond pulses. A change of~$\pi/2$ leads to the emission of an XUV continuum corresponding to an isolated attosecond pulse (c). Reprinted by permission from Macmillan Publishers Ltd: Nature Photonics ref.~\cite{NATPHOT-Krebs-2013}, copyright~(2013).}
\label{Fig2}
\end{figure*}
For the operation of HHG up to the MHz regime, the efficient conversion from the IR to the XUV spectral range requires high intensities ($\mathrm{I>10^{13} W/cm^2}$), which cannot be easily reached using directly the output of a femtosecond oscillator. A high finesse enhancement cavity offers a strategy for increasing the energy, and thus intensity, per pulse whilst maintaining the high repetition rate of the mode-locked oscillators
The coherent addition of pulses leads to intracavity average power and pulse energies on the order of several tens or hundreds of Watts and a few
$\mathrm{\mu Js}$, respectively. A gas jet for HHG is introduced in the enhancement cavity leading to harmonic generation at MHz repetition rates~\cite{NATURE-Gohle-2005, PRL-Jones-2005}.\par In the first experiments, Ti:Sa-based oscillators were used demonstrating generation of odd harmonics up to the 13th order~($\mathrm{\sim 20~eV}$) from the fundamental radiation. The harmonic radiation was extracted from the cavity
using a reflective plate oriented at the Brewster angle for the fundamental radiation. This, however, limits the performances of the enhancement cavity due to the introduction of additional dispersion and induced nonlinearities. At the same time, the low-energy per pulse offered by Ti:Sa oscillator limits the intracavity total power to about 38~W~\cite{NATURE-Gohle-2005}.\par
In order to overcome these limitations, Pupeza and coworkers introduced a new Yb-based experimental setup consisting of an 1040-nm oscillator and a chirped-pulse fibre amplifier (output power of 60~W, 78~MHz repetition rate, and 172~fs pulse duration). The pulse duration
could be further reduced to 51 fs by using an additional nonlinear compression stage based on a large mode area fibre and a chirped mirror compressor set. Such pulses were coupled in an enhancement cavity where the XUV radiation, generated in the gas jet, was coupling out of the cavity by a custom-designed pierced mirror. This output coupler reduced the intracavity losses and led to circulating power up to 5.6~kW, which resulted in a high conversion in the XUV spectral range up to a few $\mathrm{\mu W}$ for a single harmonic around 32~eV. At the same time, the coupling efficiency
increased with decreasing wavelength which favored the generation of high photon energies. Using neon as generating
medium, the harmonic cut-off was extended up to 100~eV.\par The effect of the pulse duration on the efficiency of the
intracavity HHG was investigated by this experimental setup. In general, it is well known that, for laser intensities below the ionization saturation threshold, HHG efficiency increases by decreasing the duration of the driving pulses; moreover, the use of short femtosecond pulses (down to the few-cycle regime) is a prerequisite for the generation of isolated attosecond pulses.In general, for
a fixed peak intensity, shorter pulses reduces the ionization of the gas target due to the reduced duration of the interaction. The use of short pulses is expected to improve the performances of the enhancement cavity as ionization-induced nonlinearities constitutes one
of the limiting factors for the average power circulating inside the cavity. This was experimentally demonstrated
by comparing cavity peak intensity achieved with and without gas injection for two different pulse durations (57~and~172~fs) (see Fig.~3 in ref.~\cite{NATPHOT-Pupeza-2013}).\\

Recent developments in optical parametric amplification (OPA) and optical parametric chirped-pulse amplification (OPCPA) systems based on fiber and thin-disk technology have pushed the frontier of the generation of intense, ultrashort femtosecond pulses at high-repetition rates. They are ideal driving sources for HHG and could determine a breakthrough in the generation and application of ultrashort XUV radiation~\cite{IEEE-Sansone-2012}. Krebs and coworkers have recently demonstrated the generation of XUV continua up to 600~kHz using a CEP-stable OPCPA system~\cite{NATPHOT-Krebs-2013}. The system was based on two OPA stages delivering a smooth output spectrum spanning
from 750~nm to 1,250~nm, seeded by the output of Ti:Sa CEP-stable oscillator. The two stages were pumped by a Yb-fibre laser amplifier delivering frequency-doubled 100~$\mathrm{\mu J}$, 500~fs pulses at the wavelength of 515~nm, with a repetition rate up to 600~kHz. The compressed pulse energy was 14~$\mathrm{\mu J}$ with a pulse duration of 6.6~fs. Figure~\ref{Fig2} shows the XUV spectra generated as a function of the CEP in neon where a clear CEP-dependence can be observed (Fig.~\ref{Fig2}a) which confirms the short duration of the driving pulses in the generating medium. A CEP change of $\pi/2$ drives a transition between a highly modulated spectrum (Fig.~\ref{Fig2}b) and a continuous distribution (Fig.~\ref{Fig2}c), which is a prerequisite for the generation of an isolated attosecond pulse.\par Coherent combination techniques are expected to dramatic
increase the average power of fiber-based systems in the coming years~\cite{IEEE-Limpert-2014}.
Pulses as short as 30~fs at an average power up to 163~W were achieved by combining a four-channel fiber chirped-pulse amplification system~\cite{OL-Klenke-2013} with an external compressor, based on a hollow-fiber capillary, and a set of chirped mirrors~\cite{OL-Hadrich-2013}, . Phase-matched HHG in xenon driven by this source led to the demonstration of unprecedented average power in the XUV spectral range up to $\mathrm{143 \mu W}$ at 30~eV~\cite{NATPHOT-Hadrich-2014}.

\subsection{HHG for high XUV intensities} \label{sec22}
HHG in gas targets is characterized by a low-conversion efficiency on the order, typically, of $10^{-9}-10^{-5}$, depending on the energy range and on the particular experimental conditions~\cite{NATPHOT-Sansone-2011}. Investigation of nonlinear effects on the attosecond timescale in the XUV spectral range (for example two-photon double ionization of helium~\cite{PRA-Ishikawa-2002}) requires intensities of $\mathrm{I=10^{14}-10^{15}~W/cm^2}$, corresponding to pulse energy in the $\mu$J range. This requires driving laser pulses of a few tens or hundreds of mJ for the generation and application of trains and isolated attosecond pulses for the characterization of nonlinear dynamics in atoms and molecules~\cite{PQE-Midorikawa-2008}. The first pioneering experiments focused on the demonstration of attosecond nonlinear metrology~\cite{NATURE-Tzallas-2003} by measuring the interferometric autocorrelation of an attosecond pulse train~\cite{NATURE-Tzallas-2003, PRL-Nabekawa-2006a}, which showed the $\pi$-phase flip occurring between consecutive attosecond pulses~\cite{PRL-Nabekawa-2006b}. Since then, the field has rapidly progressed towards the generation, characterization, and application of intense isolated attosecond pulses to XUV-pump-XUV-probe approaches. While trains of attosecond pulses are usually generated by multicycle driving fields, additional techniques are required for confining in time the HHG process, allowing for the creation of an isolated attosecond pulse~\cite{NATPHOT-Chini-2014}. It is important to point out that while the first demonstration of isolated attosecond pulses was based on the use of few-cycle driving fields~\cite{NATURE-Hentschel-2001}, recently the creation of isolated attosecond waveform using multicycle driving pulses was demonstrated~\cite{PRL-FENG-2009}. Tzallas and coworkers showed, for the first time, the time-resolved characterization of electronic dynamics in atoms~\cite{NATPHYS-Tzallas-2011} and molecules~\cite{PRA-Carpeggiani-2014} with a resolution down to 1~fs by implementing an XUV-pump-XUV-probe approach and using the interferometric polarization gating (IPG) technique~\cite{NATPHYS-Tzallas-2007}.\par A schematic view of the experimental apparatus used in those investigations in shown in Fig.~\ref{Fig3}a:
\begin{figure}[htb]
\centering\includegraphics[width=12cm]{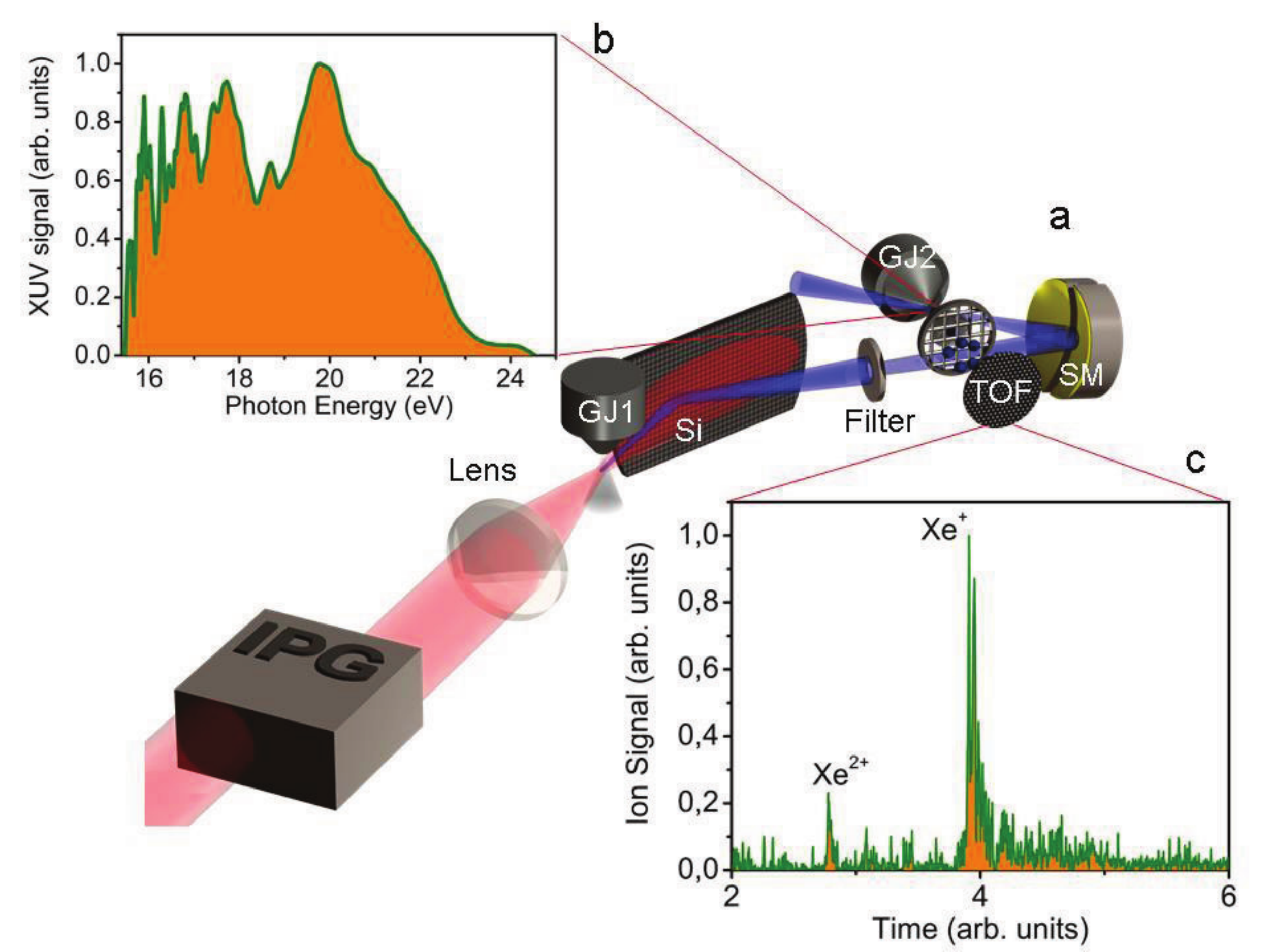}
\caption{a) Experimental setup for generation and application of intense attosecond XUV pulses. IPG: interferometric polarization gating; GJ1 and GJ2 : gas jets; SM: split mirror. b) XUV continuum generated in Xenon. c) TOF mass spectra generated by the focused XUV pulse. Reprinted by permission from Macmillan Publishers Ltd: Nature Physics ref.~\cite{NATPHYS-Tzallas-2011}, copyright~(2011).}
\label{Fig3}
\end{figure}
 the high-energy driving pulse was focused in a first gas jet~(GJ1) for efficient generation of harmonics. Loose focusing geometries are typically employed for increasing the interaction volume, and hence the total conversion efficiency. Figure~\ref{Fig3}b shows that IPG implementation results in the generation of a supercontinuum. The fundamental IR radiation was then separated by the co-propagating harmonic radiation by using a reflection off a silica plate at the Brewster angle. In this configuration, an additional metallic filter was needed in order to eliminate the residual IR radiation reflected by the plate. The attosecond pulse was focused into a second gas jet~(GJ2) by means of a split mirrors (SM) that operated also as delay stage to delay one part of the XUV beam with respect to the other~\cite{JOSAB-Kolliopoulos-2014}. The ions generated in the interaction region were finally collected by a Time-of-Flight (TOF) mass spectrometer. Due to the high-intensity of the XUV beam, nonlinear ionization occurred as shown in Fig.~\ref{Fig3}c, which reports the mass ion spectra measured in xenon. The observation of $\mathrm{Xe^{2+}}$ clearly indicates the nonlinear ionization of xenon atoms in the jet.\par Finally, the temporal structure of isolated attosecond pulses with 1.3~$\mathrm{\mu J}$ energy has been characterized using a nonlinear autocorrelation method~\cite{NATCOMM-Takahashi-2013}. The peak power of this source (2.6~GW) makes it a feasible alternative to XUV FEL for the investigation of nonlinear dynamics in the XUV spectral range.\\

An alternative approach for the generation of intense attosecond pulses is based upon HHG using solid targets. This overcomes the ionization saturation of the generating medium that limits the driving pulse energy of HHG in gas targets. In surface-based HHG, the intense driving pulse creates a plasma at the interface, which is responsible for the generation of the high-order harmonics in the direction of the reflected beam~\cite{NJP-Tsakiris-2006}. There are not any limitations to the maximum intensity of the laser field that can be applied. For relativistic intensities, the properties of the emitted radiation are independent from the target characteristics but solely depend on the properties
of the electronic plasma created by the pulse~\cite{PRE-Baeva-2006}.\par Depending on the intensity of the driving field, surface HHG can be described in terms of coherent wake emission from the plasma waves excited in the electronic plasma ($\mathrm{I<10^{18} W/cm^2}$)~\cite{PRL-Quere-2006}, or as the Doppler upshifted radiation created by the relativistic motion of the electronic plasma under the influence of the external laser field ($\mathrm{I>10^{18}~W/cm^2}$)~\cite{NATPHYS-Dromey-2006}. Harmonic radiation using the second process up to 3.8~keV was demonstrated~\cite{PRL-Dromey-2007}. The conversion efficiency associated to the process is also orders of magnitude greater than HHG in gas targets~\cite{NJP-Tsakiris-2006}. In particular, the expected conversion efficiency of a few percent in the low XUV photon energy range creates the possibility to generate mJ-level attosecond trains and isolated attosecond pulses using table-top laser system~\cite{NJP-Rykovanov-2008,PRL-Heissler-2012}.\par Temporal characterization of attosecond pulse trains generated on solid targets was recently demonstrated by implementing a second-order autocorrelation \cite{NATPHYS-Nomura-2009,NJP-Horlein-2010}. The generation and characterization of isolated attosecond pulses using this mechanism would represent a major breakthrough for nonlinear, keV attosecond spectroscopy~\cite{NJP-Tsakiris-2006}.

\subsection{HHG up to the keV spectral range} \label{sec23}
Equation~\ref{Eq1} describes the maximum photon (cut-off) energy reached in HHG in gas targets:
\begin{equation}
E=I_p+3.17U_p=I_p+3.17\frac{e^2}{8\pi^2 \varepsilon_0 m c^3} I\lambda^2
\label{Eq1}
\end{equation}
where~$I_p$ is the ionization potential, $U_p$ the ponderomotive potential, $e$ the electron charge, $m$ the electron mass, $c$ the speed of light, $I$ the pulse intensity, and $\lambda$ the field wavelength. The generation and application of trains and isolated attosecond pulses have, so far, been mainly limited to energies below~150~eV. In order to extend the cut-off energies up to the water window (280-530~eV) and to the keV range, Eq.~\ref{Eq1} suggests that an increase the intensity or the wavelength of the driving pulse is required but this formula does not incorporate saturation effects. These limit the cut-off photon energy for driving intensities exceeding the saturation intensity which is the intensity required for full ionization of the generating medium. Increasing the intensity of the pulses beyond the saturation intensity leads to the complete
ionization of the medium on the leading edge of the pulse. Under this conditions, the medium does not experience the maximum electric field intensity leading to a reduction of the cut-off photon energy with respect to the prediction of Eq.~\ref{Eq1}.
\begin{figure}[htb]
\centering\includegraphics[width=12cm]{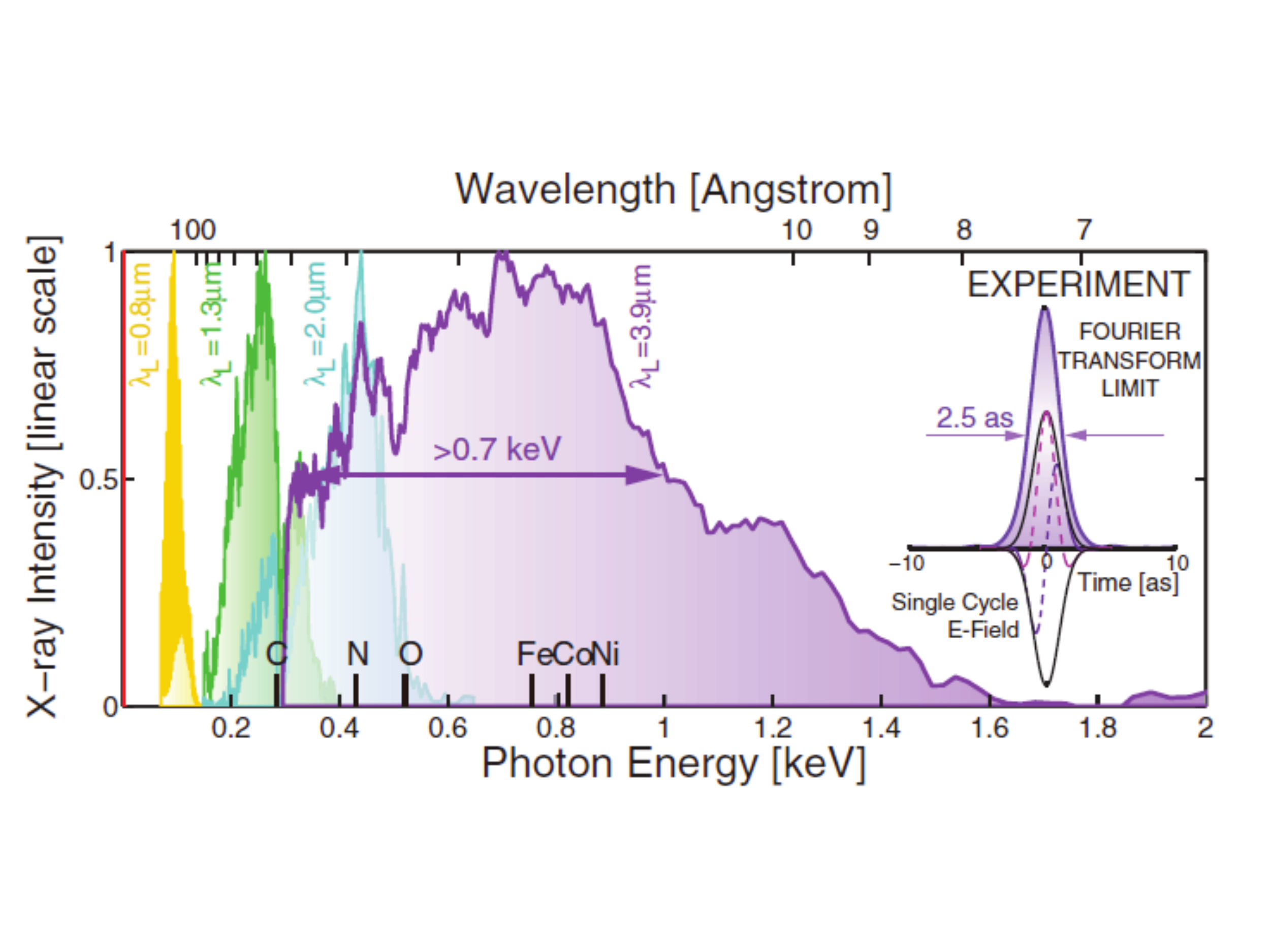}
\caption{X-ray spectra generated by a multi-mJ mid-IR driving source operating at 3.9~$\mathrm{\mu m}$. The bandwidth of the coherent radiation extends over more than 700~eV supporting a Fourier limited duration of 2.5~as. Phase-matched generation at this driving wavelength allows one to extends the cut-off energy up to 1.6~keV, much higher with respect to the cut-off achievable with shorter driving wavelengths: the XUV spectra generated by 800-nm, 1.3-$\mathrm{\mu m}$ and 2.0-$\mathrm{\mu m}$ fields are shown for comparison. From ref.~\cite{SCIENCE-Popmintchev-2012}. Reprinted with permission from AAAS.".}
\label{Fig4}
\end{figure}
The $\lambda^2$-scaling offers more advantage for the extension of the harmonic cut-off, and it has driven the development of mid-IR driving sources during the last years. Currently, driving pulses with central wavelength between 1.4 and 3.9~$\mu m$ have been demonstrated.
These sources typically exploit nonlinear effects, such as difference-frequency generation, for the generation of low-energy pulses in the mid-IR, which are then amplified in nonlinear optical parametric amplifiers. Using this approach, CEP-stable pulses can be generated~\cite{OL-Vozzi-2007} and used for harmonic generation up to 200~eV in a two-color scheme~\cite{OL-Calegari-2009} and the CEP-dependence of the HHG spectra in the water window generated by few-cycle mid-IR has recently been reported~\cite{NATCOMM-Ishii-2014}.\par
The possibility to extend the generation of coherent radiation up to the keV spectral range was recently demonstrated~\cite{SCIENCE-Popmintchev-2012}. In this work Popmintchev and coworkers used 3.9~$\mathrm{\mu m}$ with an energy of 10 mJ at a repetition rate of 20 Hz to drive the HHG process in a waveguide filled with helium at pressures as high as 35~atm. Phase-matched harmonic generation was achieved by balancing the dispersion introduced by the neutral gas and by the free-electron plasma generated by the driving field. The geometrical dispersion imposed by the guiding structure can contribute to the achievement of phase-matching condition. Figure.~\ref{Fig4} reports the experimental spectrum measured after the waveguide using a soft x-ray spectrometer and an x-ray charge coupled device camera. The X-ray spectrum extends up to 1.6 keV indicating the generation of nonlinear harmonic orders larger than 5000. The coherence properties of the emerging beam were confirmed by a Young's double slit apparatus.
In these experimental conditions, the main characteristics of the HHG process differs from IR driving laser induced HHG. Helium, despite its low polarizability, is the most efficient medium for HHG up to the keV range as the lack of core electrons prevents the re-absorption of the generated radiation and does not represent a limitation for the coherent buildup of the harmonic signal. At the same time, the high pressure required for the phase-matching condition and the long driving wavelength, determines a mean interatomic distance (about 10~$\mathrm{\r{A}}$) which is much smaller than the extension of the electronic wave function during its motion in the continuum (about 500~$\mathrm{\r{A}}$). In the case of 800-nm wavelength driving pulses, the typical extension of the electron trajectory in the continuum is 20~$\mathrm{\r{A}}$, while the interatomic separation (at 0.1~atm) is 70 $\mathrm{\r{A}}$. It can be therefore assumed that the electronic wave packet does not encounter any other atom (or ion) during its motion. This assumption is no longer valid in the case of long-driving wavelength and high pressure, but, nevertheless the scattering of the electronic wave packet does not appreciably affect the coherence properties of the X-ray beam.
The efficient generation of multicycle mid-IR pulses in waveguide is expected to lead to the confinement of the HHG process to short time windows, eventually leading to an isolated attosecond pulse~\cite{PNAS-Chen-2014}. This property is again the result of the phase-matching conditions occurring in the capillary during the propagation of the driving pulse.\par
Temporal confinement of the HHG process could lead even to the generation of zeptosecond pulses as recently predicted \cite{PRL-Garcia-2013}. The zeptosecond temporal structure arises from the interference of high harmonic emission from multiple reecounters of the photoelectron wave packet released in the continuum with the parent ion. In HHG the contributions of quantum paths revisiting more than once the original position can be neglected due to the reduced recombination probability as the result of wave packet spreading. However, on the trailing edge of the laser pulse the smaller recombination probability is compensated for by the higher ionization rate, which occurs closer to the peak of the mid-IR pulse. This leads to comparable amplitudes from trajectories that revisit possibly several times the parent ion. It is important to observe that for even longer driving wavelength ($\lambda > \mathrm{9~\mu m}$), the action of the magnetic field on the electron paths will start playing a role, introducing a new element that could affect the recombination probability with the parent ion.

\section{XUV-driven ultrafast dynamics: a few examples}
\label{sec3}
\subsection{Imaging and control of two-electron wave packet}
\label{Fano}
\begin{figure}[htb]
\centering\includegraphics[width=12cm]{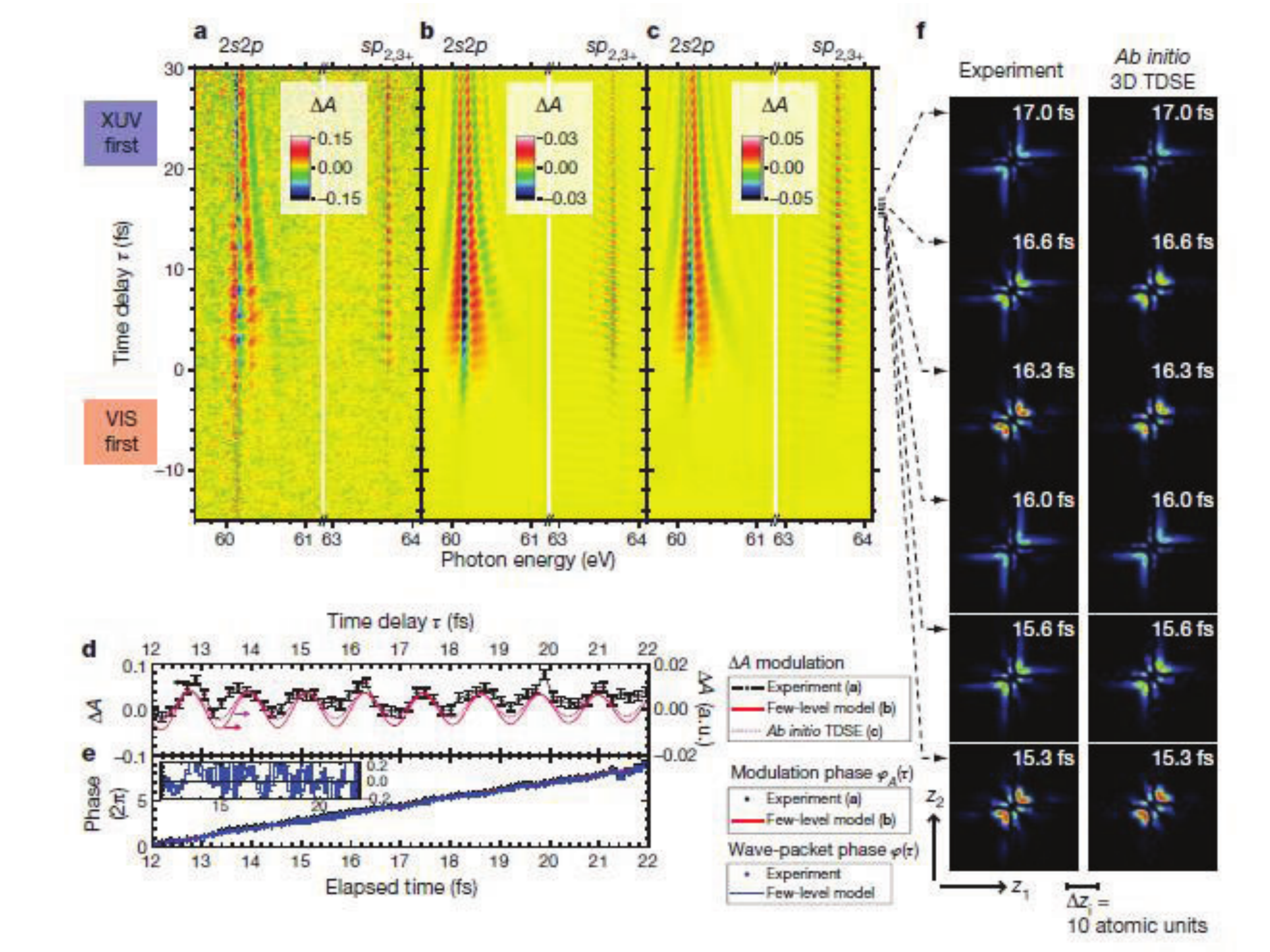}
\caption{Time-resolved observation of a two-electron wave packet in helium.
a-c: helium XUV light absorbance ($\Delta A(\omega,\tau)$) versus time delay; experiment (a), few-level model simulation (b) and ab initio calculation (c).
d: oscillation of $\Delta A(\tau)$  near $sp_{2,3+}$ resonance at 63.67~eV.
e: phase of $\Delta A(\tau)$, $\varphi_{A}(\tau)$, and relative phase $\varphi(\tau)$ of the two-electron wave packet involving the $2s2p$ and $sp_{2,3+}$ states. f: visualization of the two-electron wave-packet motion.
Left column: experimentally reconstructed wave packet, including only the two measured states $2s2p$ and $sp_{2,3+}$. Right
column: \emph{ab initio} TDSE simulation, including all excited states. Reprinted by permission from Macmillan Publishers Ltd: Nature  ref.~\cite{NATURE-OTT-2014}, copyright~(2014).}
\label{Fig5}
\end{figure}

The time-resolved observation and control of electronic dynamics driven by electron-electron correlation is one of the main goals of ultrafast XUV science, and, in particular, of attosecond physics~\cite{RMP-Krausz-2009,CHEMPHYSCHEM-Sansone-2012}.Helium is a fundamental benchmark, as it represents the simplest two-electron system and the investigation of its Fano resonances is a fundamental step towards the understanding of correlated electrons dynamics.
Fano resonances are peculiar resonant structures with an asymmetric shape which are markedly different from the common lorentzian lineshape. In atoms, Fano resonances appear in the presence of a configuration state lying above the ionization threshold. The asymmetric shape is explained as the result of the interference between two indistinguishable pathways leading to the same final state into the continuum: direct ionization or excitation and subsequent decay from the (metastable) configuration state. These resonances were first explained by Ugo Fano in 1961 in his seminal paper~\cite{PR-Fano-1961}. Several investigations since have focused upon the time-resolved observation of the resonances dynamics and have led to the time-resolved estimation of the autoionization time by measuring the time duration of the photoelectron peak corresponding to the IR-generated sideband of the resonance, as a function of the relative delay between the isolated attosecond and IR pulse~\cite{PRL-Gilbertson-2010}. The broadband attosecond pulses can lead to the excitation of several autoinizing resonances, leading to the formation of an excited two-electron wave packet that evolves on a few femtosecond or even attosecond timescale. This correlated electronic motion has been recently observed by Ott.~\emph{et al}~\cite{NATURE-OTT-2014} in a transient absorption experiment, where the observable is provided by the XUV spectrum transmitted through a thin gas cell filled with helium. A moderately intense IR pulse was used to couple the XUV-populated $2s2p$ and $sp_{2,3+}$ states via two IR photons. The difference in the phase evolution between the two components manifests itself in the periodic modulation of the transmitted XUV spectrum, as it is shown in Fig.~\ref{Fig5}. The measured data (Fig.~\ref{Fig5}a) are in excellent agreement with the two-electron wave packet evolution simulated using either \emph{ab-initio} TDSE calculations or a reduced model (Fig.~\ref{Fig5}b,c)~\cite{NATURE-OTT-2014}. From the comparison between the data and the theoretical expectation, it is possible to retrieve the relative phase between the two components of the two-electron wave packet (Fig.~\ref{Fig5}d,e) and, therefore, reconstruct the correlated electronic dynamics as shown in Fig.~\ref{Fig5}f. Only the relative phase can be reconstructed using this approach and prior knowledge of the wave function is required in order to visualize the spatial distribution of the wave packet. Nevertheless, there is excellent agreement between theoretical expectation and experiment.

\subsection{Time-resolved dynamics of Fano resonances}
\begin{figure}[htb]
\centering\includegraphics[width=12cm]{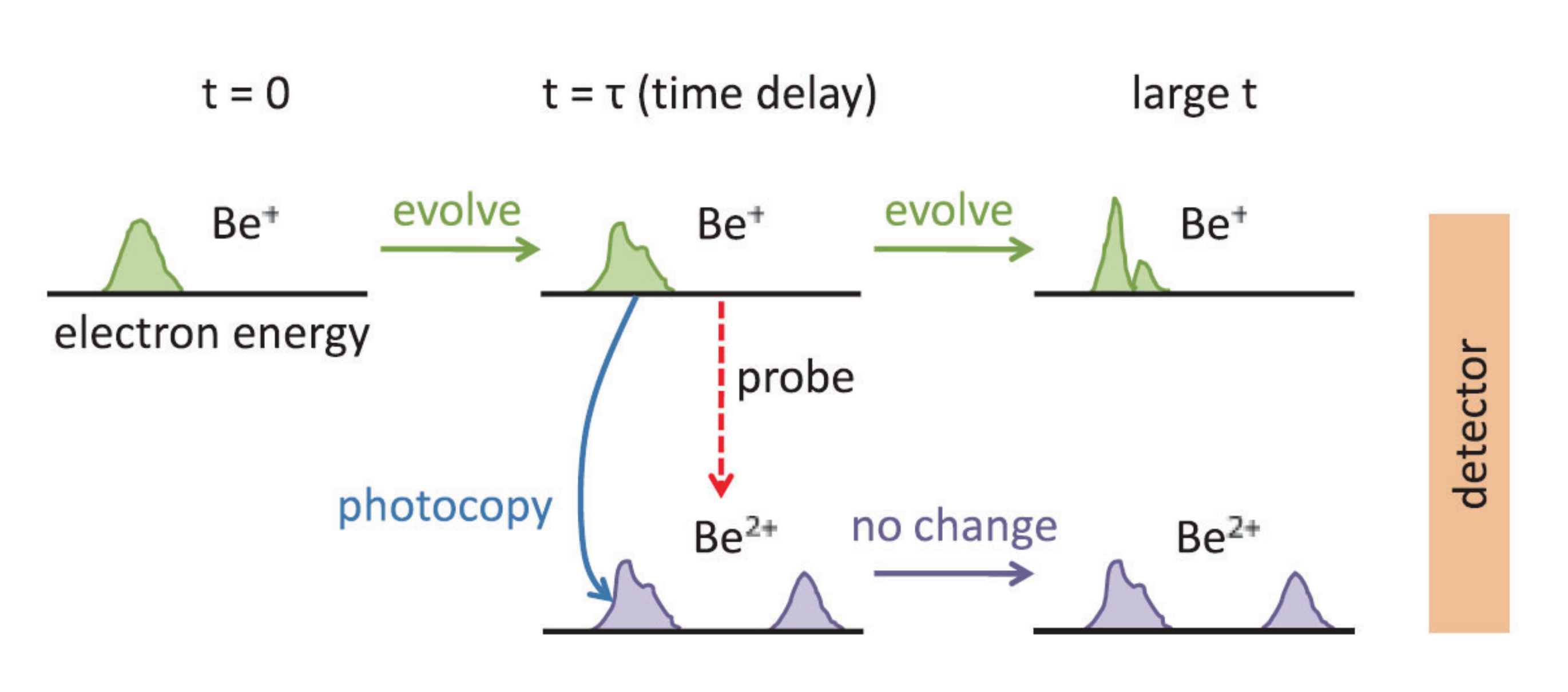}
\caption{Concept of the scheme for probing the time-dependent Fano resonance profile proposed in~\cite{PRA-Chu-2010}.  Autoionization starts at t=0, when the pump pulse is over. To determine
the evolving electron wave packet at time $\tau$, a probe pulse is used to
ionize the 2s electron, creating the $E'_p E_p$ states. After the probe is over, the new wave packet consists of a
part of the old wave packet which continues to evolve and autoionize
(upper row) and the newly generated part made of $E'_p E_p$ states
(lower row). The latter does not change with time. By collecting the energy-resolved double-ionized signals, the resonance profile in the autoionization at time $\tau$ is captured. Figure reproduced with permission from ref.~\cite{PRA-Chu-2010}, copyright~(2006) APS.}
\label{Fig6}
\end{figure}

Direct imaging of a correlated electronic wave function requires the simultaneous collection of information regarding the different degrees of freedom of the two particles, for example the momenta resulting from the ionization of the system. The time-resolved evolution of a
Fano resonance profile highlighting the dynamics associated to the formation of the characteristic lineshape was studied in ref.~\cite{PRA-Chu-2010}. In that work, an experimental scheme for the time-resolved investigation of this process was also indicated, as shown in Fig.~\ref{Fig6}. The proposed example was beryllium (electronic ground state $2s^2$), excited to the doubly excited states $2pns$ (with $n=3$ to $n=9$). These states are embedded in the $2sEp$ continuum and autoionize on a timescale of a few femtoseconds. The wave function of the system can be described as:
\begin{equation}
\psi(t)=\sum_n d_n(t)\ket{2pns}+\int d_E(t)\ket{2sEp} dE
\label{wavepacket}
\end{equation}
where $d_n(t)$ and $d_E(t)$ are time-dependent coefficient of the bound and continuum part of the electronic wave packet.
In proposed scheme, the evolution of the absorption profile can be reconstructed by using the combination of two attosecond (or few-femtosecond) pulses for exciting and probing the dynamics. The first pulse creates the wave packet described by Eq.~\ref{wavepacket} and thus starts the autoionization mechanism (see Fig.~\ref{Fig6}).
The second synchronized probe pulse projects a fraction of the wave packet $\int d_E(t)\ket{2sEp} dE$ onto states of the $\ket{E'_pE_p}$ double ionization continuum, corresponding to the ejection of the $2s$ electron and resulting in the creation of a doubly charged ion $\mathrm{Be^{2+}}$. Since these are stationary states of the dication, the coefficients describing their evolution, $d_{E'E}(\tau)$, depend only on the delay $\tau$ and do not evolve further in time (differently from $d_n(t)$ and $d_{E}(t)$, because of the autoionization process). Thus, sampling the observable $d_{E'E}(\tau)$ at different delays allows for the reconstruction of time-resolved absorption profile, which, for large delays, converges to the expected Fano shape~\cite{PRA-Chu-2010}.\\
It is necessary to measure in coincidence the two electrons or one electron and the double charged ion $\mathrm{Be^{2+}}$ in order to characterize the $d_{E'E}(\tau)$ function. This experimental approach challenges current experimental possibilities in attosecond science as it requires intense XUV sources (a XUV-pump-XUV-probe approach needs to be implemented) at high repetition rate to generate statistically meaningful data. Suitable
attosecond sources are currently not available but these experiments will become feasible with newly developed large scale facilities XUV photon sources such as FELs and ELI-ALPS.

\subsection{Time-resolved dynamics of Auger processes}
The interaction of an XUV or X-ray pulse (from several tens of eV up to a few keV) can lead to the formation of highly charged ions, through a mechanism initiated by emission of an inner-valence or core-shell electron leaving the ion in an excited state. The initial hole can be filled through an Auger decay process leading to the emission of a second electron and the formation of doubly charged ions. Depending on the external photon flux, a second photon can be eventually absorbed from the same (or a different) inner-valence or core-shell triggering a second Auger decay process, and so on. As a result the ionization potential increases at each step until it exceeds the XUV/X-ray photon energy. At this point, further ionization can only occur by emission of valence electrons. For intense X-ray pulses in the keV, the interplay between these mechanisms was investigated in ref.~\cite{NATURE-Young-2010}.

The core-shell ionization process can also occur according to a non-sequential mechanism for pulse durations comparable to the Auger decay lifetime. In this case, the second XUV (X-ray) photon can be absorbed when the first hole has not yet been filled (before the first Auger decay), leading to the formation of a double core hole. These states decay with the emission of an electron whose energy is displaced with respect to the energy of the single Auger decay. This gives a very specific fingerprint process which could be used for chemical analysis~\cite{PNAS-Berrah-2011}. Moreover, X-ray fluorescence from a double core hole state was recently used to infer information about the duration of an intense X-ray pulse in the few-femtosecond domain~\cite{PRL-Tamasaku-2013}.\par
Time-resolved observation of Auger decay on the few-femtosecond timescale was first reported in ref.~\cite{NATURE-Drescher-2002} where core-level ionization was triggered by the absorption of a single photon from an isolated attosecond pulse and the lifetime of of the relaxation process was determined with an analogous technique used in the determination of Fano resonance lifetime~\cite{PRL-Gilbertson-2010} (see section~\ref{Fano}). In the XUV/IR pump probe experiment sidebands of the main Auger-electron line are created and the duration of the sidebands as a function of the relative delay between the IR and XUV pulses allows one to estimate the decay time of the core-hole state.\par
\begin{figure}[htb]
\centering\includegraphics[width=10cm]{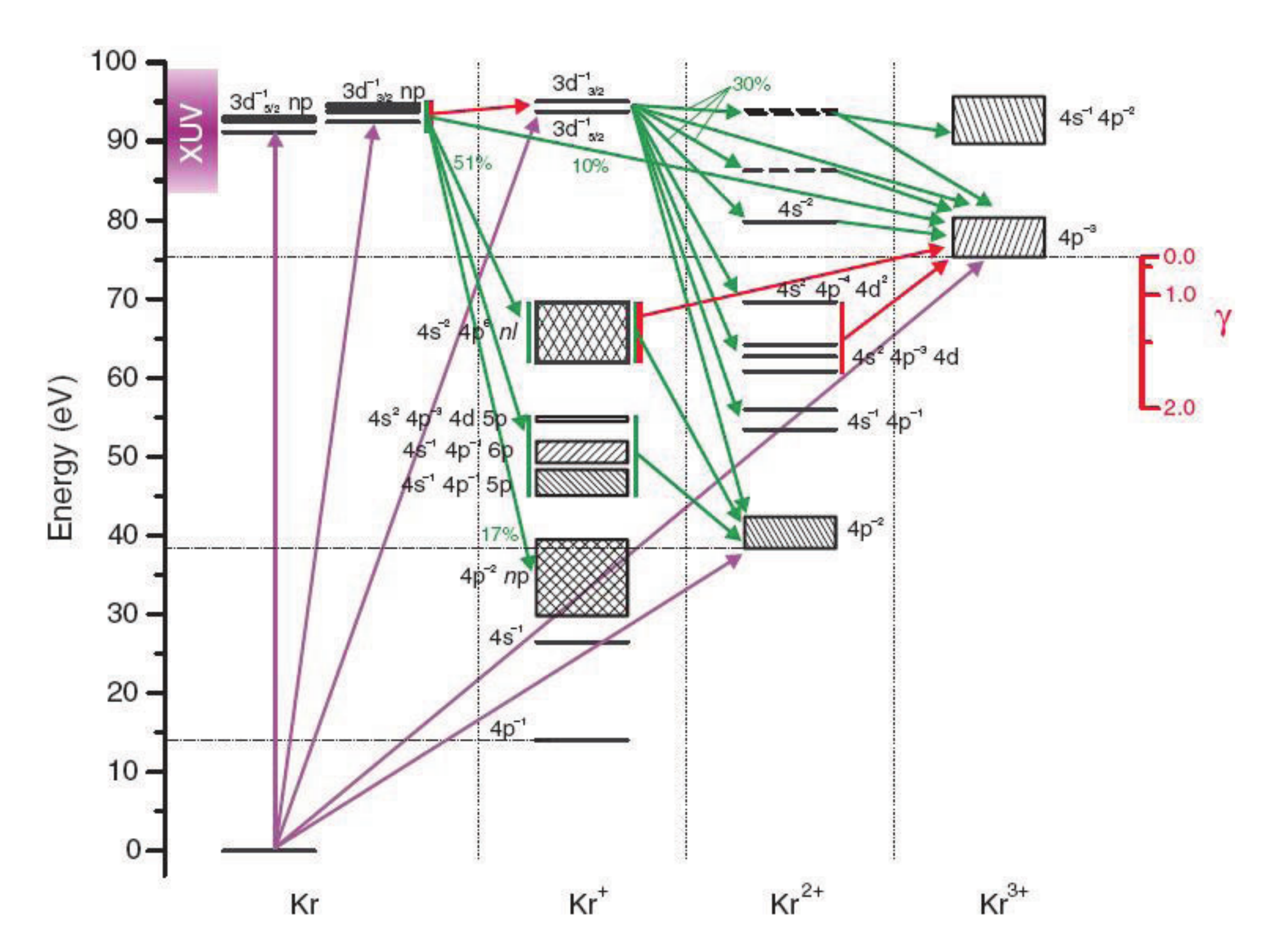}
\caption{Schematic representation of energy levels and transitions in neutral,
singly, doubly and triply ionized krypton. After the $3d$ excitation by the attosecond XUV pulse (magenta arrows), different relaxation channels (green arrows) open up: resonant and normal Auger decay as well as cascaded Auger decay. Figure reproduced with permission from ref.~\cite{NJP-Uphues-2008}, copyright(2008) IOP.}
\label{Fig7}
\end{figure}
\begin{figure}[htb]
\centering\includegraphics[width=10cm]{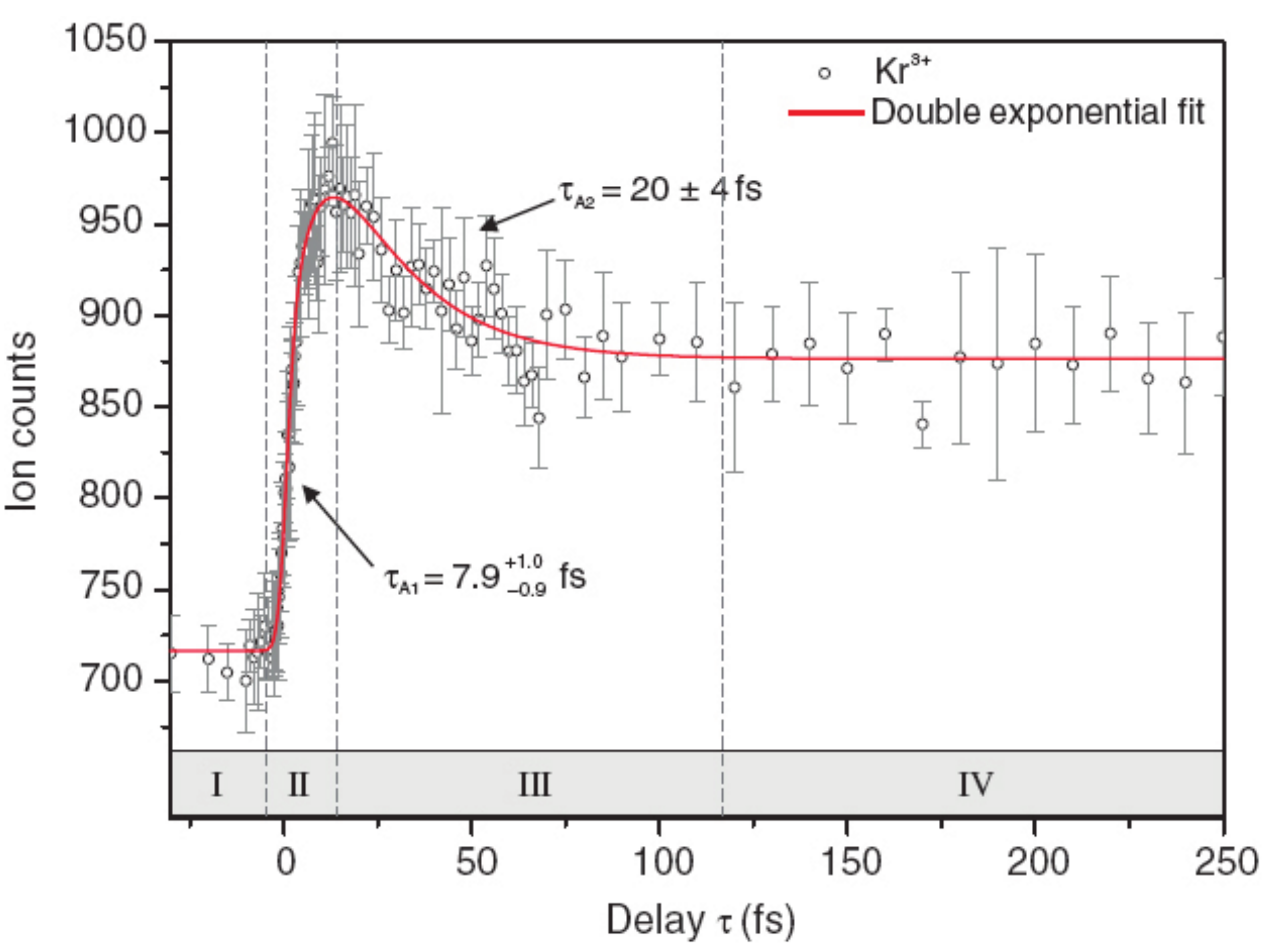}
\caption{Ion chronoscopy determination of the lifetime of Auger decay in krypton. $\mathrm{Kr^{3+}}$ signal is plotted against the XUV/IR delay. The rise time allows for the estimation of the lifetime, since the formation of $\mathrm{Kr^{3+}}$ ions is made possible by the building up (and subsequent IR-ionization) of the $\mathrm{Kr^{2+}}$ population resulting from the Auger process. Figure reproduced with permission from ref.~\cite{NJP-Uphues-2008}, copyright(2008) IOP.}
\label{Fig8}
\end{figure}

An alternative approach based on the measurement of the ion-charge state and yield was demonstrated in ref.~\cite{NATURE-Uiberacker-2007}, by monitoring the ion yield for different charge states as a function of the relative delay. The scheme can be understood considering Fig.~\ref{Fig7}: after the Auger decay process, $\mathrm{Kr^{3+}}$ can be populated by IR-ionization of $\mathrm{Kr^{2+}}$ ion states. The formation of these ionic states can occur only after the Auger decay process and presents a rise-time that reflects the autoionization dynamics. In the measurement shown in Fig.~\ref{Fig8}, the increase of the signal presents a rise-time of $\tau_{A1}=7.9$~fs, which is consistent with the value obtained from the sidebands duration~\cite{NJP-Uphues-2008}.\par
HHG-based sources offer also the possibility to time-resolve the cascaded Auger decay process.
Indeed, after the first Auger decay, a sequential second decay can occur which leads to the emission of a third electron. The associated dynamics were probed in time by measuring the time evolution of xenon ion yield as a function of the relative delay between the initial attosecond pulses and a delayed few-cycle infrared pulse~\cite{NATURE-Uiberacker-2007}.
While ion chronoscopy reveals integral information regarding the cascaded Auger dynamics, photoelectron spectroscopy allows the assignment of intermediate states and resonances in the process. This approach was exploited in~\cite{NJP-Verhoef-2011} to investigate the resonant and normal Auger decay in Krypton excited by an isolated attosecond pulse centered at 94~eV. The XUV-pump/IR-probe time-resolved measurement indicated a lifetime of about 70~fs in the relaxation dynamics of the excited ionic states in the cascaded resonant Auger decay. Comparison with theoretical models led to the attribution of this lifetime to the second-step Auger decay of the resonantly excited $3d^{-1}np$ states with $n=6,7$.\\

\section{The Extreme Light Infrastructure Attosecond Light Pulse Source}
\label{sec4}
The Attosecond Light Pulse Source (ELI-ALPS) is one of the three pillars of the Extreme Light Infrastructure (ELI), a new EU research initiative that has cutting edge, future oriented light based fundamental science and technology as its primary mission. ELI-ALPS will provide some of the shortest and most powerful light pulses with the highest repetition frequency, all within XUV/X-ray wavelength range. The short duration of the individual
XUV/X-ray light pulses, $<$ 1~fs and the intrinsic attosecond synchronization with other light fields ranging from the THz over the IR/mid-IR into the visible, and bursts of elementary particles (electrons and ions) distinguishes ELI-ALPS from alternative high-brightness X-ray sources such as FELs and accelerators. To the scientist ELI-ALPS will offer an unique possibility to explore non-linear processes at and beyond the XUV, to observe and control ultrafast electronic and structural dynamics of atoms, molecules, clusters, liquids and solids, and to investigate new ideas for biological research, material science and medical applications. ELI-ALPS will maintain state-of-the-art end stations including reaction microscope (REMI), cold-target recoil ion spectrometer (COLTRIMS), electron/ion 3D momentum imaging and ion microscopy, velocity map imaging (VMI), angle-resolved photoemission spectrometer and photoelectron emission microscope. In addition, there will be the possibility of installing specialized apparatus.\par
\begin{figure}[htb]
\centering\includegraphics[width=10cm]{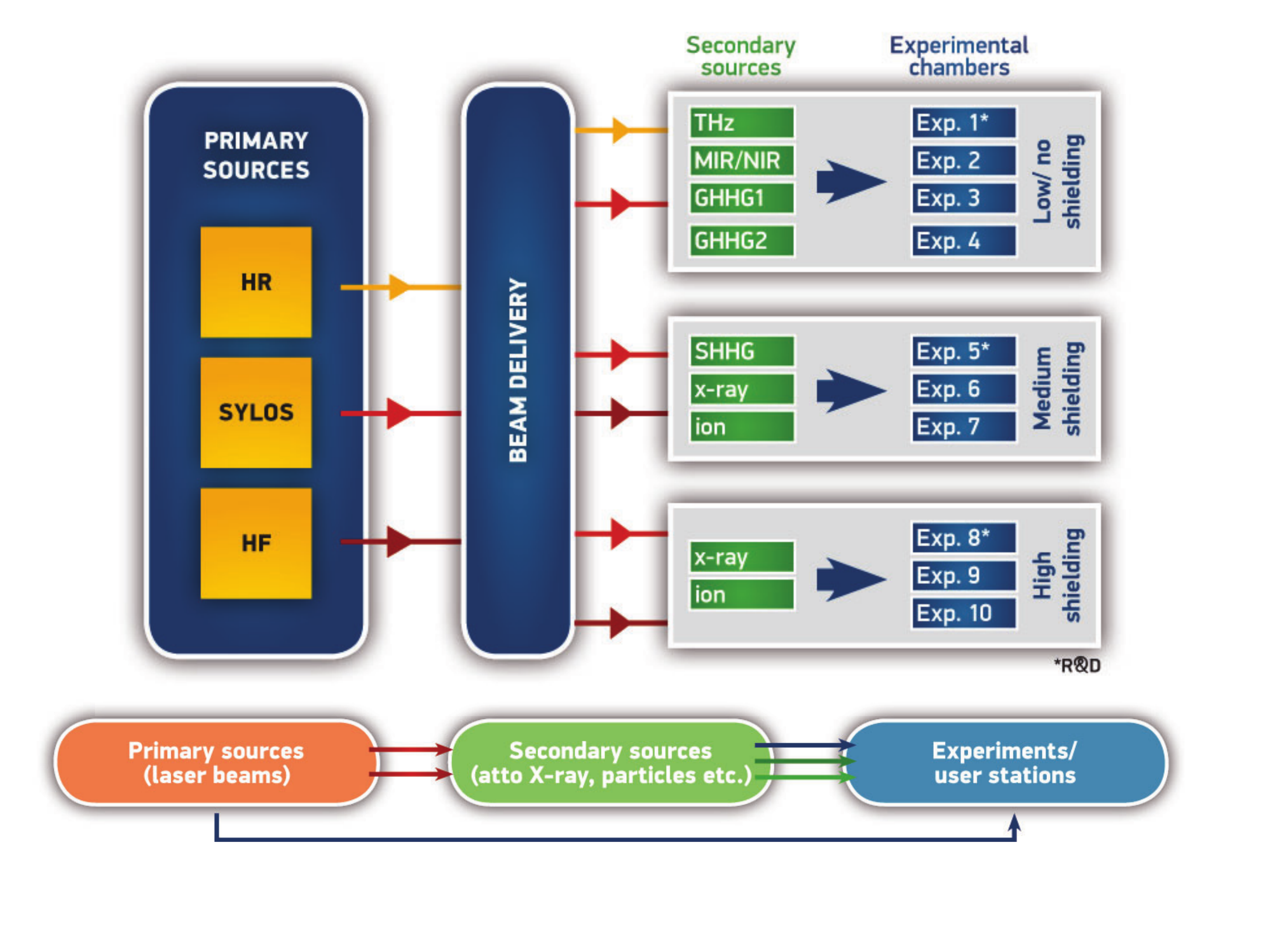}
\caption{Schematics of the research equipment of the ELI-ALPS facility. The primary sources will be delivered to several beamlines, where secondary sources consisting in attosecond pulses, X-ray radiation, THz pulses, and particle beams are generated. These sources will finally drive experiments and user end stations in three different areas accordingly to the shielding characteristics required (Low-, medium-, and high-shielded target areas)}
\label{Fig9}
\end{figure}

 The research equipment of ELI-ALPS is schematically shown in Fig.~\ref{Fig9}. It will feature four main driving laser sources (primary sources), which will drive XUV and X-ray photon beamlines and particles (electron and ions) beamlines (secondary sources). Different combination of primary and secondary sources will be delivered to experimental end stations for time-resolved experiments (experiments/user stations).\par Table~\ref{table1}
summarizes the main characteristics of the primary sources. Due to the challenging specifications, which are well-beyond the state-of-the-art, the implementation of these sources has been planned in two different phases (Phase I and Phase II). The characteristics for Phase I have already been consolidated whilst the expected, final performances for the second implementation phase depends on the outcome of the first phase.\par The high-repetition rate (HR) primary source will operate at 100~kHz and will drive two XUV photon beamlines based on HHG in gas targets (GHHG) These secondary sources will provide users with trains and isolated attosecond pulses for time-resolved experiments. The expected performances for the XUV photons secondary sources are summarized in Table~\ref{table2}.\par The single cycle laser (SYLOS) will favor energy at the expenses of repetition rate, operating at 1~kHz for a pulse energy of 100~mJ. This primary source will drive two GHHG beamlines, one beamline for the generation of harmonics from solid targets (SHHG beamlines), and one beamline for the generation and acceleration of electron beam to be used in pump-probe experiments in combination with other secondary or primary beams.\par Finally, the high-field (HF) source will deliver pulses with 2~PW peak-power for the investigation of plasma physics and for exploring the potential of surface harmonic generation through a dedicated SHHG beamline (x-ray).\\

The availability of light pulses with duration well below 1~fs has opened the possibility to study the pure electronic dynamics of valence electrons in atoms and molecules. In analogy to the familiar creation of vibrational wave packets in femtospectroscopy, attosecond valence electron spectroscopy aims at creating electronic wave packets on a time scale that is considerably faster than relaxation or nuclear motion in molecules. The evolution of such wave packets can lead to localization of charge at specific locations of the molecules and thus open chemical reaction pathways that would not be favored in a conventional chemical reaction~\cite{NATPHOT-Lepine-2014}. Understanding the dynamics of such wave packets is subject to recent attosecond studies. Experimentally, the problem is approached by either two color pump-probe schemes, where one pulse is of attosecond duration while the other may be femtoseconds long, or directly by attosecond pump-probe schemes consisting of identical XUV replica. In the former case, the challenge is not so much the pulse energy but rather the requirement for the tunability of the photon energies and bandwidths, a stable, well-defined and adjustable attosecond and femtosecond time structure, and the precise synchronization of the two-color field. An IR pulse is frequently used to probe the excitation either by attosecond streaking or by coupling distinct quantum paths in the ionization and dissociation process. Obviously, the synchronization with the attosecond pulse is crucial for the temporal resolution, while the IR/mid-IR photon energy determines the time range that can be accessed or the states which are coupled. All these requirements can be satisfied by the all-optical approach envisaged at ELI-ALPS. Shorter than 5~fs CEP stabilized IR pulses are promised at mJ pulse energies from which state-of-the-art high harmonic pulses in the few 10~nJ range and mid-IR pulses in the $\mathrm{10~\mu J}$ range are derived. What is more, this operation is performed at high repetition rates of up to 100~kHz. This unique feature is indispensable not only to achieve an unprecedented signal quality and thus statistically meaningful data, but will also enable access to higher order, multi-parameter detection schemes~\cite{JPB-Rudenko-2010}.\\

An illustrative example for the capabilities of ELI-ALPS is the non-linear interaction of atoms with XUV pulses, introduced in section~\ref{sec2}B. The two-photon double ionization of helium is the most fundamental realization of an attosecond pump-probe experiment~\cite{PRA-Ishikawa-2002}. The theory of this rich and involved three-body problem has matured but still awaits experimental testing~\cite{JPB-Palacios-2010}. Thorough understanding of this system is essential for the design and the interpretation of experiments on more complex systems. Due to the stringent requirements on the pulse parameters~\cite{JPB-Nikolopoulos-2001} it has only recently become possible to detect $\mathrm{He^{2+}}$ from a two-photon process induced by high-order harmonics~\cite{PRL-Nabekawa-2005}. To date there is still no full kinematic data available owing to the very low reaction yield and the low repetition rate (typically less than 10~Hz) of the driving pulses available so far. In this experiment pulses of 10~fs and 24~nJ were used with $\mathrm{I_{peak}=10^{13} W/cm^2}$. For a comprehensive investigation of such processes ELI-ALPS will eventually generate sub-femtosecond pulses reaching intensities well beyond $\mathrm{10^{14}~W/cm^2}$. This will be the starting point for the thorough understanding of the dynamics and interferences of the correlated electrons predicted for delayed pulse experiments~\cite{JPB-Palacios-2010,PRA-Lambropoulos-2008} as a benchmark for all further experiments.\\

\begin{table}

\caption{\label{table1} Main parameters of the primary sources that will be available at ELI-ALPS.}
\begin{minipage}{\textwidth}
\scalebox{0.7}{
\begin{tabular}{lccccc}
Name&Central wavelength (nm)&Energy (mJ)&Rep. rate (kHz)& Number of cycles & CEP (pulse-to-pulse) (mrad)\\ \hline
HR (Phase I) & 1030 & 1 & 100 & 1.85 & 100 	\\
HR (Phase II)\footnote{Correspond to expected parameters \label{PrSoa}}& 1030 & 5 & 100 & 1.8 & 100 		\\
Mid-IR & 3100 & 0.15 & 100 & 4 & 100 	\\
SYLOS (Phase I) & 800 & 45 & 1 & $<$3.7 & 250 	\\
SYLOS (Phase II)\textsuperscript{\ref{PrSoa}} & 800 & 100 & 1 & $<$2 & 250 		\\
HF & 800 & 34000 & 0.01 & 6.5 & 250 	\\
\end{tabular}
}
\end{minipage}

\end{table}

Typical photon energies for the excitation of small molecules such as $\mathrm{H_2}$ are around 15-35~eV, while the separation of subsequent excited electronic levels is on the order of 3-20~eV corresponding to 100-600~as pulses in the XUV. Even broader phase-locked XUV spectra are required if continuum states are to be included in the wave packet. A recent example that would substantially benefit from ELI-ALPS's sources is the asymmetry of $\mathrm{D^{+}}$ ejection from two-color ionization of $\mathrm{D_2}$ that results from the steered electron localization by a timed XUV/IR field~\cite{PRL-Fischer-2010, NATURE-Sansone-2010}. It may be regarded as the original demonstration on how light-induced electron migration influences the outcome of a chemical reaction through the coupling of electronic and nuclear degrees of freedom. Related experiments have been undertaken by utilizing a XUV/XUV pump-probe scheme where the attosecond probe beam ionizes the evolving molecule at a preset delay time yielding detectable fragments~\cite{PRA-Carpeggiani-2014}. In contrast to the XUV/IR scheme this approach is fully perturbative and insights into the intrinsic charge dynamics are obtained. The time scales for the simultaneous but clearly distinct electronic and nuclear wave packet motion reaches into the sub 1~fs region~\cite{PNAS-Palacios-2014}.\par
The interplay between the electronic and nuclear degrees of freedom beyond the Born-Oppenheimer regime is a largely unexplored but highly important topic~\cite{PhysScripta-Muskatel-2009}. In larger molecules, the migration of the hole or electron wave packet following photoionization or photoexcitation can be probed in a similar manner and the yield of specific fragments indicates the charge distribution at the moment of detection~\cite{SCIENCE-Calegari-2014}. In this way the time-dependent redistribution of the electronic density can be inferred. These processes occur on the time scale of 1-10~fs~\cite{ChemPhys-Kuleff-2007}. Just as for the homo-nuclear molecules, correlated detection schemes and simultaneous acquisition of several observables will open new insights into the intramolecular charge dynamics of these larger molecules.\par
The high repetition rates at high to very high photon energies available at ELI-ALPS will also allow to venture into the study of charge transfer processes~\cite{NATURE-Fohlisch-2005} and the dynamics of molecular orbitals that occur at the interfaces of solids. In some cases reorganization times well below 1~fs are foreseen, while in others no experimental data exists yet. Further examples include light-induced charge migrations and charge density waves (e.g. $\mathrm{TaS_2}$)~\cite{OSA-Cacho-2012} and phase transitions in dielectrics (e.g. SiO2)~\cite{NATURE-Schiffrin-2013} or carrier dynamics in graphene~\cite{NATMAT-Gierz-2013}. Novel techniques may also make solid state phenomena accessible to attosecond spectroscopy~\cite{CHEMPHYS-Ivanov-2013}.\\

What has been said about valence electron dynamics can directly be transferred to the time-resolved spectroscopy of core electron excitations. For these studies ELI-ALPS provides X-ray sources up to 10 keV similar to those in the XUV. The dynamics following removal of a core electron are decidedly more complex than for valence excitations but can be resolved in two- or one-color pump-probe schemes as has been demonstrated for I2 and $\mathrm{C_2H_2}$, respectively. Optically highly ionized atoms may be counted into this category~\cite{JPB-Schippers-2014}. Experimental shortcomings due to timing jitter, pulse duration and pulse energy can easily be overcome by ELI-ALPS's all-optical approach. The ultrafast relaxation of highly excited molecular states has important implications for the radiation resistance of biomolecules. An understanding of these mechanisms is still in its infancy but electronic correlations and proton transfer on the fs time scale appear as major factors.\\

\begin{table}
\caption{\label{table2} Main parameters of the XUV photon secondary sources that will be available at ELI-ALPS}
\begin{minipage}{\textwidth}
\scalebox{0.7}{
\begin{tabular}{lcccc}
Type & Photon energy range (eV)& Energy ($\mathrm{\mu J}$)& Pulse duration (as)&  Synchronization\footnote{Synchronization between attosecond and IR pulses (GHHG HR) and between two replica of the intense attosecond pulse (GHHG SYLOS)} (as)\\ \hline
GHHG HR (Phase I) & 17-90 & $2\cdot10^{-4}-5\cdot10^{-6}$ & $<$ 600 &  $<$900\\
GHHG HR (Phase II)& 17-90 & $3\cdot10^{-4}-7\cdot10^{-6}$ & $<$ 600 & $<$900\\
GHHG SYLOS (Phase I) & 17-100 & 0.1-0.001 & $<$ 800 & $<$30 	\\
GHHG SYLOS (Phase II)& 17-100 & 0.5-0.005 & $<$ 1500 & $<$30\\
SHHG SYLOS (Phase II)& 10-400 & 100-0.1 & -\footnote{Due to the novelty of SHHG in this regime, no predictions can be made. The parameter will be subject of R\&D activity \label{SeSoa}} & -\textsuperscript{\ref{SeSoa}}	\\
SHHG HF & 10-1,000 & 1000-5& -\textsuperscript{\ref{SeSoa}} & -\textsuperscript{\ref{SeSoa}} \\
\end{tabular}
}
\end{minipage}
\end{table}
A highly demanding non-linear process is the two-photon ionization of inner shells. Estimates show that ELI-ALPS will be the first of its kind capable to provide these. The promise is the ability to study fundamental inner shell dynamics with the attosecond time resolution facilitated by optical pulses. Another utilization of core electrons is the holographic imaging of the nuclear configuration with a photoelectron ejected from the molecule itself. This method also relies on the energy selectivity and the timing accuracy attainable with ELI-ALPS's secondary sources.\par
In the experiments discussed so far information about the dynamics of the electronic density is obtained only indirectly by watching specific parts of a molecule after cleavage or from spectral changes. This requires some knowledge of the nuclear and electronic structure at the time of observation. Spectroscopy is insensitive to electronic configuration changes that do not involve measurable differences in energy. A direct probing technique for electronic distribution dynamics is time-resolved X-ray diffraction and imaging techniques which are well established in femtosecond spectroscopy. It is clear that for the understanding of light-matter interactions and correlated motions to extend such measurements to the attosecond time scale with a spatial resolution that is compatible with molecular and atomic dimensions. This approach is known as 4-dimensional imaging. ELI-ALPS will be equipped to fill this need by offering reproducible X-ray pulses of 100 nJ down to $\mathrm{1-12~\mathrm{\r{A}}}$ along with synchronized attosecond pulses to initiate the dynamics of interest. Experiments planned at ELI-ALPS will include the light-field induced charge dynamics in ordered dielectrics and crystals~\cite{NATURE-Schultze-2013,NATURE-Schiffrin-2013}, and photonic nanostructures at optical frequencies. Recent studies reveal that coherent motions of the coupled electronic-nuclear system play a hitherto underestimated role in the charge transfer of organic molecules~\cite{SCIENCE-Falke-2014}. Spatial tracking of photo-initiated electron and proton motion in natural and artificial molecules (e.g. light harvesting complexes or photochromic molecules) is another area of broad scientific and technological interest. Their ultrafast time scales and their complex structure make the 4-dimensional approach indispensable. Dynamics in small gas-phase molecules~\cite{PRL-Kupper-2014}, larger functional molecules~\cite{SCIENCE-Calegari-2014}, molecules in the liquid-phase~\cite{SS-Brown-2013, PRL-Nordlund-2007}, plasma formation~\cite{PRL-Liseykina-2013} and plasma wave evolution~\cite{PRL-Varin-2012} in atomic nano-clusters, and metallic nanostructures~\cite{NATPHOT-Stockman-2007, NATURE-Aeschlimann-2007} are further examples that make the light based approach a convincing step.
ELI-ALPS will also ambitiously pursue the extension of this method towards shorter wavelengths on the order of 0.1~$\mathrm{\r{A}}$ and to true 3-dimensonal spatial imaging. This will allow to follow the charge density dynamics of core electrons and to obtain the full, unambiguous set of coordinates for the electronic and nuclear positions in time.

\section{Acknowledgments}
Financial support by the Alexander von Humboldt Foundation (Project "Tirinto"), the Italian Ministry of Research (Project FIRB  No. RBID08CRXK). This project has received funding from the European Union$\mathrm{'}$s Horizon 2020 research and innovation
programme under the Marie Sklodowska-Curie grant agreement No. 641789 "MEDEA" (Molecular Electron Dynamics investigated by IntensE Fields and Attosecond Pulses) and the Hungarian Scientific Research Fund (OTKA project NN 107235) . We thank Prof.~Dimitris~Charalambidis for fruitful discussions and continuous support.

e-mail: giuseppe.sansone@polimi.it\\

\section*{References}

\bibliography{Rev_bib}

\end{document}